\documentclass[twocolumn]{aastex63}
\usepackage{latexsym,amssymb,amsbsy,amsmath,natbib}
\usepackage{array}
\defcitealias{Scoville_2016}{S16}
\defcitealias{Planck11}{Planck Collaboration 2011}

\DeclareGraphicsExtensions{.pdf,.png,.jpg}

\shorttitle{Evolution of molecular hydrogen density}
\shortauthors{T. K. Garratt et al.}

\begin{document}

\title{Cosmic evolution of the H$_2$ mass density and the epoch of
  molecular gas}

\author{T.~K.~Garratt}\email{t.garratt@herts.ac.uk}
\affil{Centre for Astrophysics Research, University of Hertfordshire, Hatfield, AL10 9AB, UK.}
\author{K. E. K. Coppin}
\affil{Centre for Astrophysics Research, University of Hertfordshire, Hatfield, AL10 9AB, UK.}
\author{J. E. Geach}
\affil{Centre for Astrophysics Research, University of Hertfordshire, Hatfield, AL10 9AB, UK.}
\author{O. Almaini}
\affil{School of Physics and Astronomy, University of Nottingham, University Park, Nottingham, NG7 2RD, UK}
\author{W. G. Hartley}
\affil{Department of Astronomy, University of Geneva, ch. d'\'Ecogia 16, CH-1290 Versoix, Switzerland}
\author{D. T. Maltby}
\affil{School of Physics and Astronomy, University of Nottingham, University Park, Nottingham, NG7 2RD, UK}
\author{C. J. Simpson}
\affil{Gemini Observatory, Hilo, HI 96720, USA}
\author{A. Wilkinson}
\affil{Sterrenkundig Observatorium, Universiteit Gent, Krijgsl
aan 281 S9, 9000 Gent, Belgium}
\author{C. J. Conselice}
\affil{School of Physics and Astronomy, University of Nottingham, University Park, Nottingham, NG7 2RD, UK}
\author{M. Franco}
\affil{Centre for Astrophysics Research, University of Hertfordshire, Hatfield, AL10 9AB, UK.}
\author{R. J. Ivison}
\affil{European Southern Observatory, Karl Schwarzschild Strasse 2, Garching, Germany}
\author{M. P. Koprowski}
\affil{Institute of Astronomy, Faculty of Physics, Astronomy and Informatics, Nicolaus Copernicus University, Grudziadzka 5, 87-100 Torun, Poland}
\author{C. C. Lovell}
\affil{Centre for Astrophysics Research, University of Hertfordshire, Hatfield, AL10 9AB, UK.}
\author{A. Pope}
\affil{Department of Astronomy, University of Massachusetts, 710 North Pleasant Street Amherst, MA 01003, USA}
\author{D. Scott}
\affil{Dept. of Physics \& Astronomy, University of British Columbia, Vancouver, Canada}
\author{P. van der Werf}
\affil{Leiden Observatory, Leiden University, P.O. Box 9513, NL-2300 RA Leiden, The Netherlands}

\begin{abstract}

We present new empirical constraints on the evolution of $\rho_{\rm  H_2}$, the cosmological mass density of molecular hydrogen, back to $z\approx2.5$.  We employ a statistical approach measuring the average observed $850\,\mu$m flux density of near-infrared selected galaxies as a function of redshift.  The redshift range considered corresponds to a span where the $850\,\mu$m band probes the Rayleigh-Jeans tail of thermal dust emission in the rest-frame, and can therefore be used as an estimate of the mass of the interstellar medium (ISM). Our sample comprises of ${\approx}150,000$ galaxies in the UKIDSS-UDS field with near-infrared magnitudes $K_{\rm AB}\leq25$\,mag and photometric redshifts with corresponding probability distribution functions derived from deep $12$-band photometry.  With a sample approximately $2$ orders of magnitude larger than in previous works we significantly reduce statistical uncertainties on $\rho_{\rm H_2}$ to $z\approx2.5$.  Our measurements are in broad agreement with recent direct estimates from blank field molecular gas surveys, finding that the epoch of molecular gas coincides with the peak epoch of star formation with $\rho_{\rm  H_2}\approx2\times10^7\,{\rm M_\odot}\,{\rm Mpc^{-3}}$ at $z\approx2$.   We demonstrate that $\rho_{\rm H_2}$ can be broadly modelled by inverting the star-formation rate density with a fixed or weakly evolving star-formation efficiency.  This ``constant efficiency'' model shows a similar evolution to our statistically derived $\rho_{\rm H_2}$, indicating that the dominant factor driving the peak star formation history at $z\approx2$ is a larger supply of molecular gas in galaxies rather than a significant evolution of the star-formation rate efficiency within individual galaxies.
\end{abstract}


\section{Introduction}

Three intimately linked observational tracers broadly characterise the cosmic evolution of galaxies: the volume averaged star-formation rate density $\rho_{\dot M_\star}(z)$, the stellar mass density $\rho_{M_\star}(z)$, and the molecular gas density $\rho_{M_{\rm H_2}}(z)$.  Our current understanding of galaxy evolution is largely driven by comprehensive measurements of the former two \citep[see][for a review]{madau14}, with a clear empirical picture emerging of an evolution of star formation, which rises rapidly to a peak around $z\approx2$ and then decays to the present day. 
Completing the triptych is important since the evolution of the molecular gas content of galaxies encodes several important pieces of astrophysics: gas consumption in star formation; gas recycling via feedback; and fresh gas accretion. Ultimately, it is the evolution of molecular gas that drives galaxy evolution as it is the fuel from which stars are assembled.  Measurements of molecular gas in galaxies are therefore needed to complete the picture, and to resolve a key outstanding question: \textit{Was the peak of star formation history driven by a larger supply of molecular gas or because galaxies formed stars more efficiently (e.g., driven by galaxy mergers/instabilities etc.), or both?}

The bulk of the cold gas reservoir in the Universe is comprised of hydrogen gas in the form of atomic hydrogen (H$_{\rm I}$) and molecular hydrogen (H$_2$).  In the current model of galaxy formation gas is delivered into galaxies via hot- or cold-mode accretion \citep[e.g.,][]{birnboim03}.  The cooling gas must form H$_2$ for star formation to occur. The two main routes to H$_2$ formation in galaxies are via the gas phase reaction ${\rm H+e^- \rightarrow H^- + \gamma}$, ${\rm H^- +  H \rightarrow H_2 + e^-}$, and via a dust phase, where H$_2$ forms on the surface of dust grains via efficient three-body reactions \citep{gould1963}.  

H$_2$ radiates poorly in typical ISM conditions due to the lack of a permanent dipole moment and a minimum rotational excitation temperature that is significantly higher (${\approx}500$\,K) than typical temperatures of the cold star-forming ISM \citep{wakelam}.  However, H$_2$ can be indirectly traced through its interactions with CO, which traces the same cold, dense ISM and has a low dipole moment enabling its excitation in regions of low density ($n_{\rm crit} \sim 10^2\,{\rm cm}^{-3}$).  Consequently, $^{12}$CO, the second most abundant molecule in the ISM, is commonly used as a tracer of the available reservoir of molecular gas in galaxies \citep[e.g.,][]{Solomon05,carilli13}.  The ground state transition CO\,($J=1\rightarrow0$) is a reliable tracer of total molecular gas, with the conversion factor from CO luminosity to H$_2$ gas mass ($\alpha_{\rm CO} = M_{\rm {H}_2}/L'_{\rm CO}$) calibrated locally \citep[see][for a review]{bolatto13}.  Observing the ground-state transition line avoids additional uncertainties inherent in observations of higher-J CO transitions, which require a correction for gas excitation to derive the equivalent CO\,($J=1\rightarrow0$) luminosity.  

Until recently measurements of the cosmological molecular gas mass density were hampered by a paucity of observational data.  Over the past few years direct measurements of the cold molecular gas reservoirs of individual galaxies have increased rapidly, with surveys primarily targeting star-forming and lensed galaxies \citep[e.g.,][]{frayer98,frayer99,coppin07,tacconi10,ivison11,thomson12,bothwell13,riechers13,stach17,oteo18,gomez19,lenki19}. However, as these surveys rely on observationally-expensive detections of faint spectral lines, measurements of molecular gas mass are still dwarfed in number in comparison to the samples for which star-formation rate (SFR) and stellar mass estimates are derived.  Moreover, to properly assess the cosmological evolution of the cold gas content of galaxies a blank field survey approach is required to measure the gas mass function, rather than targeted (and therefore biased) observations of high-{\it z} galaxies as has generally been the case for cold gas observations outside the local volume.

Recently surveys using a blank field molecular line scan strategy have emerged as an alternative to targeted observations.  These surveys evade many of the biases towards massive star-forming galaxies inherent in targeted approaches. The inaugural blank field CO survey employed the Plateau de Bure Interferometer in observations of the \textit{Hubble} Deep Field North \citep{walter14,decarli14}.  This was followed more recently by the ALMA Spectroscopic Survey in the \textit{Hubble} Ultra Deep Field \citep[ASPECS;][]{walter16,decarli16b,decarli16,aravena16b,aravena16a,bouwens16a,carilli16}, the ASPECS Large Program \citep[ASPECS LP;][]{Decarli19,gonzalez19,boogaard19,popping19,Decarli20} and the CO Luminosity Density at High Redshift survey \citep[COLDZ;][]{pavesi18,riechers19}.  These surveys have presented results setting out valuable new constraints on the evolution of $\rho_{\rm H_2}$ over a redshift range $0\lesssim z \lesssim 7$, obtained through blank field observations of CO line emission.  However, due to low number statistics and the small survey areas (which are prone to strong clustering-enhanced sample variance) used to derive the CO luminosity functions, these measurements are hampered by large statistical uncertainties.

To combat the shortfall in direct measurements of molecular gas \citet{Scoville_13,Scoville_14,Scoville_2016,Scoville_2017} employed a complementary approach that utilises submillimeter observations of the long wavelength dust continuum as a measure of the molecular gas mass in galaxies.  The Rayleigh-Jeans (RJ) tail is nearly always optically thin, and consequently measurements of dust emission can be used as a direct probe of molecular gas mass \citep[e.g.,][]{eales12,magdis12}.  Whilst ordinarily a conversion from dust to gas mass would require dust emissivity and dust-to-gas abundance to be constrained, \citet{Scoville_14,Scoville_2016,Scoville_2017} circumvent this by deriving an empirically calibrated \textit{RJ luminosity-to-gas mass} ratio using CO\,($J=1\rightarrow0$) and submillimeter continuum observations of a sample of normal star-forming and starburst galaxies at low-$z$ and submillimeter galaxies (SMGs) at high-$z$.  This approach requires assumptions about dust temperature and the evolution of the gas-to-dust mass ratio, but provides molecular gas mass ($M_{\rm mol}$) estimates within factor of ${\approx}2$ accuracy \citep[e.g.,][]{Scoville_2016,kaasinen19}. Since dust continuum measurements can be made in minutes \citep[in contrast to CO line observations which can take multiple hours, e.g.,][]{bothwell13,Tacconi13} this method can be used to derive molecular gas measurements for much larger samples of galaxies.

The \citet{Scoville_2016} RJ luminosity-to-gas mass calibration has been used to estimate the molecular gas mass for $\sim700$ ALMA-detected galaxies from the COSMOS field, with \citet{Scoville_2017} deriving molecular gas masses for individual galaxies at redshifts $0.3<z<4.5$ and \citet{liu2019} extending this approach to redshifts of $z\approx6$.  This method has also been used in combination with stacking methodologies to estimate average molecular gas masses for large samples of galaxies.  \citet{Millard20} use the \citet{Scoville_2016} calibration and apply this to a sample of $63,658$ galaxies to derive the gas mass fraction out to $z\approx5$. \citet{magnelli20} use a method similar to \citet[][and which is equivalent at solar metallicity]{Scoville_2016} and apply this to a sample of $555$ galaxies to derive the molecular gas mass density to $z\approx3$.

In this paper we contribute to the picture of cosmic galaxy evolution by building on the approach of \citet{Scoville_13,Scoville_14,Scoville_2016,Scoville_2017} estimating the evolution of the cosmological mass density of molecular hydrogen to $z\approx 2.5$ via the average submillimeter continuum emission of a sample of $150{,}000$ galaxies selected from a deep near-infrared survey in the well studied UKIDSS-UDS field. The Ultra-Deep Survey (UDS) is the deepest component of the UK InfraRed Telescope (UKIRT) Infrared Deep Sky Survey \citep[UKIDSS;][]{lawrence07}. We limit our estimate of the molecular gas mass density to $z\approx2.5$ as the \citet{Scoville_2016} calibration has only been shown to be robust out to this redshift.  Adopting a statistical approach allows us to take advantage of a near-infrared selected sample which is an order of magnitude larger than in surveys that measure dust emission \citep[e.g.,][]{Scoville_2017,liu2019,magnelli20} or CO spectral line emission \citep[e.g.,][]{walter16,riechers19,Decarli19,kaasinen19} for individual sources.  Our method differs from previous stacking approaches \citep[e.g.,][]{Millard20,magnelli20} as we do not use a combination of spectroscopic and photometric redshifts for our binning.  Instead, in the absence of a sample complete with spectroscopic redshifts, we utilise the full photometric redshift probability distribution functions for all our sources.  Our method is complementary to previous works in this field \citep[e.g.,][]{Decarli20,liu2019,magnelli20,riechers19} and allows us to reduce the statistical uncertainties on the cosmological molecular gas mass density out to $z\approx2.5$.  

This paper is organized as follows: in Section $2$, we define the maps and catalogs used; in Section $3$, we present a $3$-dimensional stacking method which we employ to measure the average (stacked) observed $850\,\mu$m flux densities for near-infrared selected galaxies as a function of redshift; in Section $4$, we show that the approach of \citet{Scoville_2016} can be applied to our stacked $850\,\mu$m flux densities to derive the cosmological molecular gas density to $z\approx2.5$.  We also demonstrate that the cosmic molecular gas density can be broadly modelled by $2$ complementary approaches (i) from the halo mass function assuming a constant halo mass range, and employing stellar-halo mass and ISM-stellar mass ratios, and (ii) inverting the star-formation rate density assuming a ``constant efficiency'' model, and in Section $5$ we interpret the overall evolution of the cosmic molecular gas mass density in the context of our results and in comparison to previous works. We present our conclusions in Section $6$.  We assume a {\textit Planck} $2015$ cosmology, where $\Omega_{\rm m}= 0.31,\, \Omega_{\Lambda}= 0.69,\, H_0= 68\, {\rm km\,s^{-1}\,Mpc^{-1}}$ \citep{Planck15}, and a \citet{Chabrier_2003} Initial Mass Function.  The AB magnitude system is used throughout. 

\section{Data}

\subsection{SCUBA-2 Cosmology Legacy Survey}
The UKIDSS-UDS field was mapped at $850\,\mu$m as part of the Submillimetre Common-User Bolometer Array 2 (SCUBA-2) Cosmology Legacy Survey \citep{geach17}. The full details of the data collection, reduction and map properties are given in \citet{geach17}.  Briefly, the beam-convolved map spans approximately $1\,{\rm {deg}}^2$ covering the bulk of the multi-wavelength coverage of this field, with a uniform (instrumental) noise of $\sigma_{850}=0.9$\,mJy\,beam$^{-1}$. \citet{geach17} estimate the SCUBA-2 confusion limit to be $\sigma_{\rm conf}=0.8$\,mJy\,beam$^{-1}$. The beam full width half maximum (FWHM) is approximately 15$''$, with a full analytic
description of the point spread function (PSF) given by \citet{geach17}.

\subsection{UKIDSS-UDS ultraviolet--optical--mid-infrared imaging and catalog}

The UDS Data Release 11 (DR$11$) $12$-band matched catalog is \textit{K}-band selected with the $95\%$ completeness limit estimated to be $K_{\rm AB}=25$\,mag. The full details of this catalog will be comprehensively provided in Almaini et al., (in prep.) and Hartley et al., (in prep.), and only a summary is given here.  The catalog provides photometry in 12 bands (\textit{U, B, V, R, $i'$, $z'$, Y, J, H, K,} $3.6\,\mu$m and $4.5\,\mu$m), where available.  

 The \textit{J, H,} and \textit{K} photometry is taken from the DR11 release of UKIDSS-UDS.  The UKIDSS project, described in \citet{lawrence07} utilises the UKIRT Wide Field CAMera \citep[WFCAM; ][]{casali07}. The photometric system and calibration are outlined in \citet{hewett06} and \citet{hodgkin09}, respectively, and the pipeline processing and science archive are described in Irwin et al., (in prep.) and \citet{hambley08}.  UKIDSS-UDS covers an area of $0.8\,{\rm deg}^2$, reaching median depths of \textit{J}$=25.6$, \textit{H}$=25.1$, and \textit{K}$=25.3$ ($5 \sigma$, {\sc AB},  estimated from $2''$ apertures in source free areas; Almaini et al. in prep.).

 The \textit{B, V, R, $i'$,} and \textit{$z'$} optical imaging is from the Subaru/\textit{XMM-Newton} Deep Survey, which utilises Suprime-Cam on the Subaru Telescope \citep{Furusawa08}.  \textit{U}-band data are from the Canada-France-Hawaii Telescope Megacam instrument (Almaini et al. in prep.) and \textit{Y}-band imaging is obtained from the VISTA Deep Extragalactic Observations survey \citep[VISTA-VIDEO:][]{jarvis13}. The InfraRed Array Camera (IRAC) imaging at $3.6\,\mu$m and $4.5\,\mu$m is from the Spitzer UKIDSS Ultra Deep Survey (SpUDS: PI Dunlop), combined with deeper data from the Spitzer Extended Deep Survey \citep[SEDS:][]{ashby13}.  To expand the coverage to outer regions of the field, shallower data are also used from the SIRTF Wide-area InfraRed Extragalactic survey \citep[SWIRE:][]{lonsdale}. 

  UDS DR11 provides image masks, with masked regions corresponding to image boundaries, artefacts, and bright stars. We employ the UDS binary mask for ``good'' regions which has an unmasked area of $0.64\,{\rm deg}^2$.  This binary mask combines the masked regions of the photometry images detailed above (not including the deeper SEDS or SpUDS IRAC images). 
  
  We also utilise the subsets feature of the UKIDSS-UDS catalog and for our galaxy sample chose the catalog-defined ``good galaxy'' subset, which comprises $217{,}429$ sources.  These sources have full $12$-band photometry and lie within the corresponding ``good'' mask regions, are not cross-talk sources (for which \textit{JHK} photometry is likely compromised), and are not classified as stars.  
  
  UDS DR11 also includes photometric redshifts derived using the code {\sc eazy} \citep[Easy and Accurate $Z_{\rm phot}$ from Yale; ][]{brammer08}.  To estimate the photometric redshift for each source the 12-band broadband photometry was fit with a spectral energy distribution template producing a redshift probability distribution (Hartley et al. in prep.).  We utilise both the maximum-likelihood photometric redshifts and redshift probability distributions provided with UDS DR11.  {\sc eazy} performs well compared to other commonly used photometric redshift codes \citep[e.g., {\sc zphot}, {\sc HyperZ}, {\sc Rainbow};][]{dahlen13}, with the resulting normalized mean absolute deviation between {\sc eazy} derived photometric redhifts and spectroscopic redshifts found to be only $\sigma_{\rm nmad}\approx 0.02$ (Hartley et al. in prep.).   

\section{Methods}\label{sect:method}

We employ a $3$-dimensional stacking approach based on the simultaneous stacking algorithm {\sc simstack} \citep[presented in detail in][]{Viero13}.  This method allows for the simultaneous fitting of the average observed flux density for multiple populations that contribute to the flux density in the observed map (such as a population of galaxies split into bins of redshift).  Importantly, this method takes into account the (usually) large beam in single-dish submillimeter maps with simulations demonstrating that this method returns an unbiased estimate of the average observed flux density for beam sizes ranging from ${\rm FWHM} = 15$--$35''$ \citep{Viero13}.  This approach also mitigates against boosting of stacking signals from clustered galaxies \citep[e.g.,][]{chary10,alberts14}. 

Our goal is to find the average observed submillimeter flux densities at given redshift intervals for a population of near-infrared selected galaxies that best fit the observed flux density in the SCUBA-2 map, taking into account the convolution of point sources with the large beam.  In this work, rather than binning galaxies by discrete photometric redshift values, we split our sample across redshift intervals according to the redshift probability distribution of each source.

First, we define our sample, performing a selection in observed $K$-band total magnitude, $K_{\rm AB}\leq25$\,mag, with the faint-end corresponding to the $95\%$ completeness limit of the UKIDSS-UDS catalog, giving us a sample of 153,399 galaxies.  At this limiting magnitude the $95\%$ stellar mass completeness is $\approx10^{9.5}\,{\rm M}_\odot$ at $z=2.5$ (Wilkinson et al. in prep.).
The redshift probability distribution, $\mathcal{P}(z)$, for each source is discretized in bins of $\Delta z$ (Hartley et al. in prep.). We make a completeness correction to the redshift probability distribution of each source, such that $\mathcal{P}(z)$ of a source of magnitude $K$ integrates to $C(K)^{-1}$, where $C(K)$ is the catalog completeness at $K$ (Hartley et al. in prep.). We assume there is no systematic redshift bias in $C(K)$ for this correction.

With the sample defined we consider a sky model in which each galaxy contributes a flux density that can be described as
\begin{align}
S_\nu = \int_0^\infty \mathcal{S}_\nu(z)\mathcal{P}(z){\rm d}z.
\label{eq sky model}
\end{align}
where $\mathcal{P}(z)$ is the normalized redshift probability distribution function and $\mathcal{S}_\nu(z)$ is the flux density
``weighting'' at redshift $z$. In practice we have discrete redshift probability distributions defined over $R$ bins such that, for a population of $N$ galaxies, the flux density in the $ij$th pixel of a map can be written
\begin{align}
M_{ij} = \sum^{N_{ij}}_p\sum^R_q \mathcal{S}_{\nu}(z_q)_{p,ij}
\mathcal{P}(z_{q})_{p,ij} \Delta z 
\label{eq Mij}
\end{align}
Because of the PSF, the flux contribution of each galaxy is distributed over many pixels according to the convolution
\begin{align}
\mathcal{M} = M\otimes{\rm PSF}.
\label{eq conv M}
\end{align}
In effect equation \ref{eq sky model} uses $\mathcal{P}(z)$ to split each of the $N$ galaxies in our $K$-selected sample into $R$ redshift bins and assumes that the galaxies in each redshift bin can be represented by an average observed flux density, $\langle S_\nu(z) \rangle$. This is effectively the ``stacked'' flux density. 

With the model sky defined we consider an optimization problem where the set of average observed flux densities, $\langle S_\nu(z) \rangle$, per redshift interval in equation \ref{eq sky model} are unknown coefficients. A key decision in defining our sky model is in the binning of $\mathcal{P}(z)$. The UKIDSS-UDS $\mathcal{P}(z)$ are binned in non-linear steps of $1 + z_{(n+1)} = 1.001 (1 + z_n)$.  This would result in hundreds of free parameters across the redshift range of interest, which is computationally impractical as well as unnecessary given the photometric redshift uncertainties.  Instead we bin each $\mathcal{P}(z)$ to $\Delta z=0.5$, giving $20$ equally-sized bins across the redshift range $0<z\lesssim10$.  

\begin{figure*}
\figurenum{1a}
\includegraphics[width = 1\textwidth]{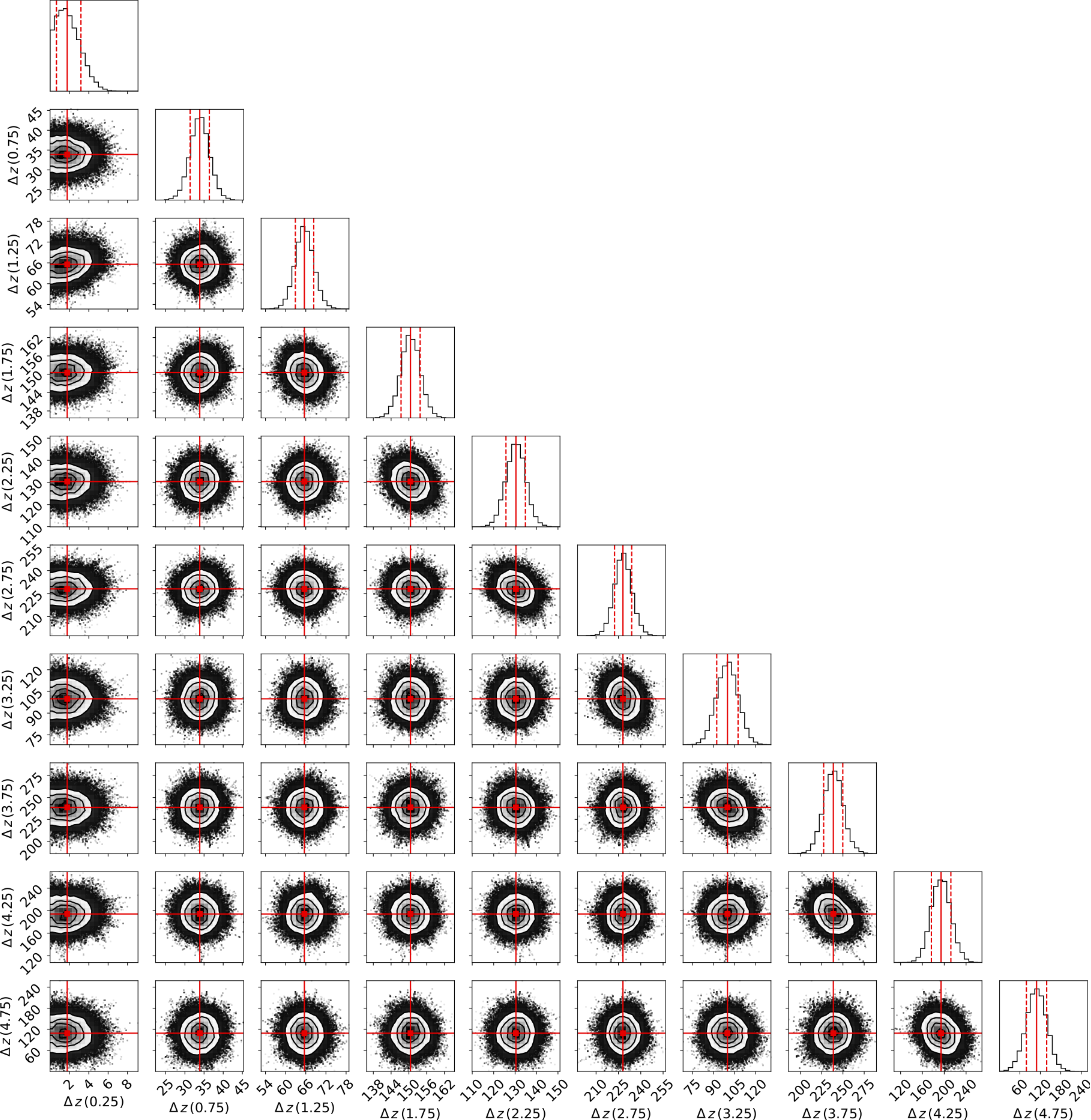}
\caption{Standard \texttt{emcee} corner plot showing the one- and two-dimensional posterior distributions for our parameters (the average observed flux density in $\mu$Jy of galaxies in each redshift bin) for the redshift intervals $\Delta z\,(0.25)$--$\Delta{z}\,(4.75)$. The density of the points and the contours correlate with the posterior probability distributions from a $1000$-step run (with $500$ steps for burn-in discarded) based on our sky model and SCUBA-$2$ maps, and employing the ``delete one'' jackknife technique with the map segment area corresponding to $A=1$ deleted to take into account sample variance \citep{Tukey58}.  The vertical red lines show the average $850\,\mu$m flux density of galaxies for each redshift interval, with the dashed red lines showing the associated $1\sigma$ uncertainties (these values are not corrected for the influence of the CMB, see Table \ref{table:results1} for CMB corrected-estimates).  As evidenced by this plot there are only very weak correlations between our parameters for the redshifts intervals $\Delta z\,(0.25)$--$\Delta{z}\,(4.75)$.}
\label{fig:emcee_a}
\end{figure*}

\begin{figure*}
\figurenum{1b}
\includegraphics[width = 1\textwidth]{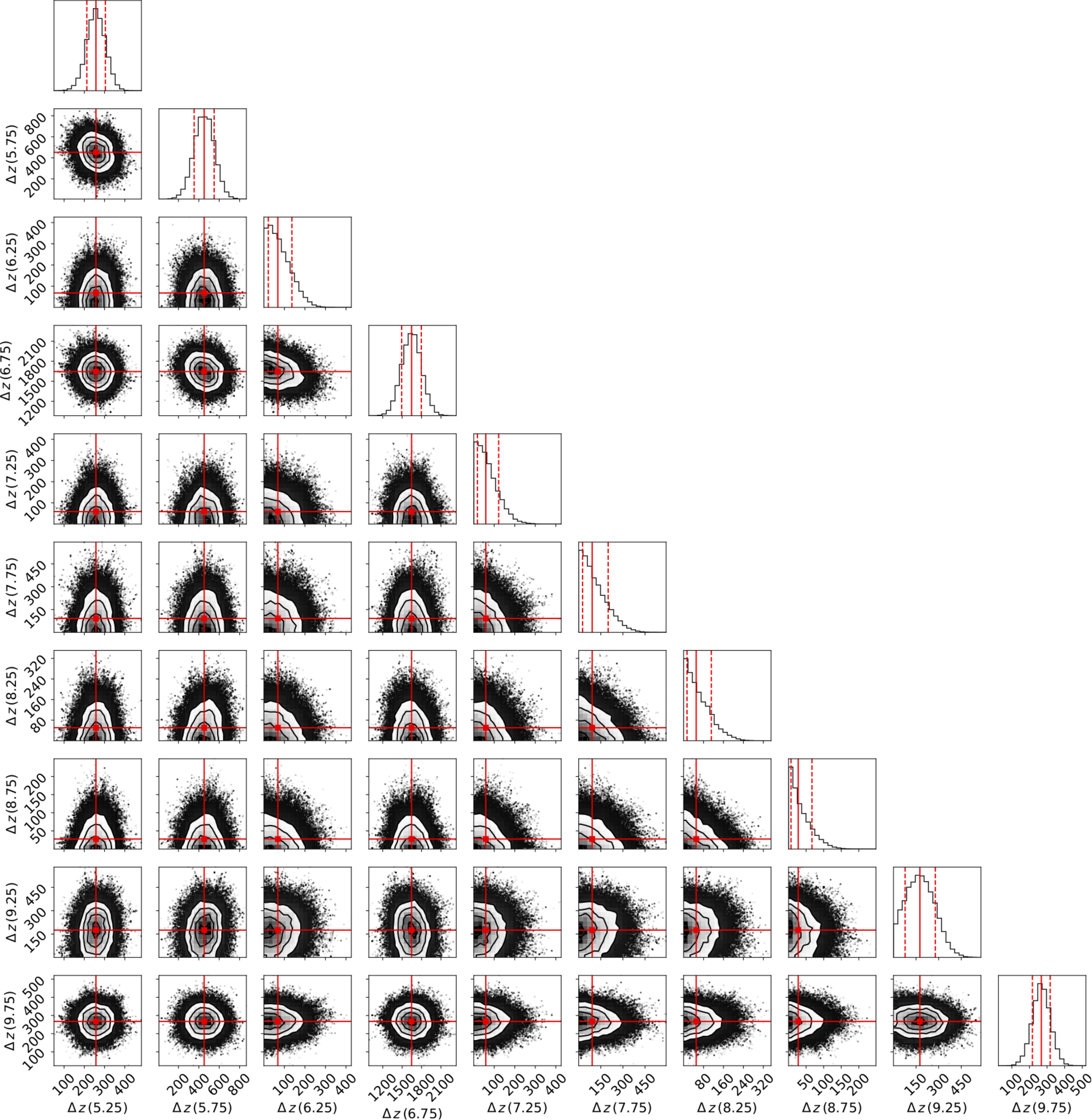}
\caption{Standard \texttt{emcee} corner plot showing the one- and two-dimensional posterior distributions for our parameters (the average observed flux density in $\mu$Jy of galaxies in each redshift bin) in the redshift intervals $\Delta z\,(5.25)$--$\Delta{z}\,(9.75)$. Detailed description as in Figure \ref{fig:emcee_a}.  As evidenced by this plot there are only very weak correlations between our parameters for the redshifts intervals $\Delta z\,(5.25)$--$\Delta{z}\,(9.75)$.}
\label{fig:emcee_b}
\end{figure*}

\begin{figure*}
\figurenum{1c}
\includegraphics[width = 1\textwidth]{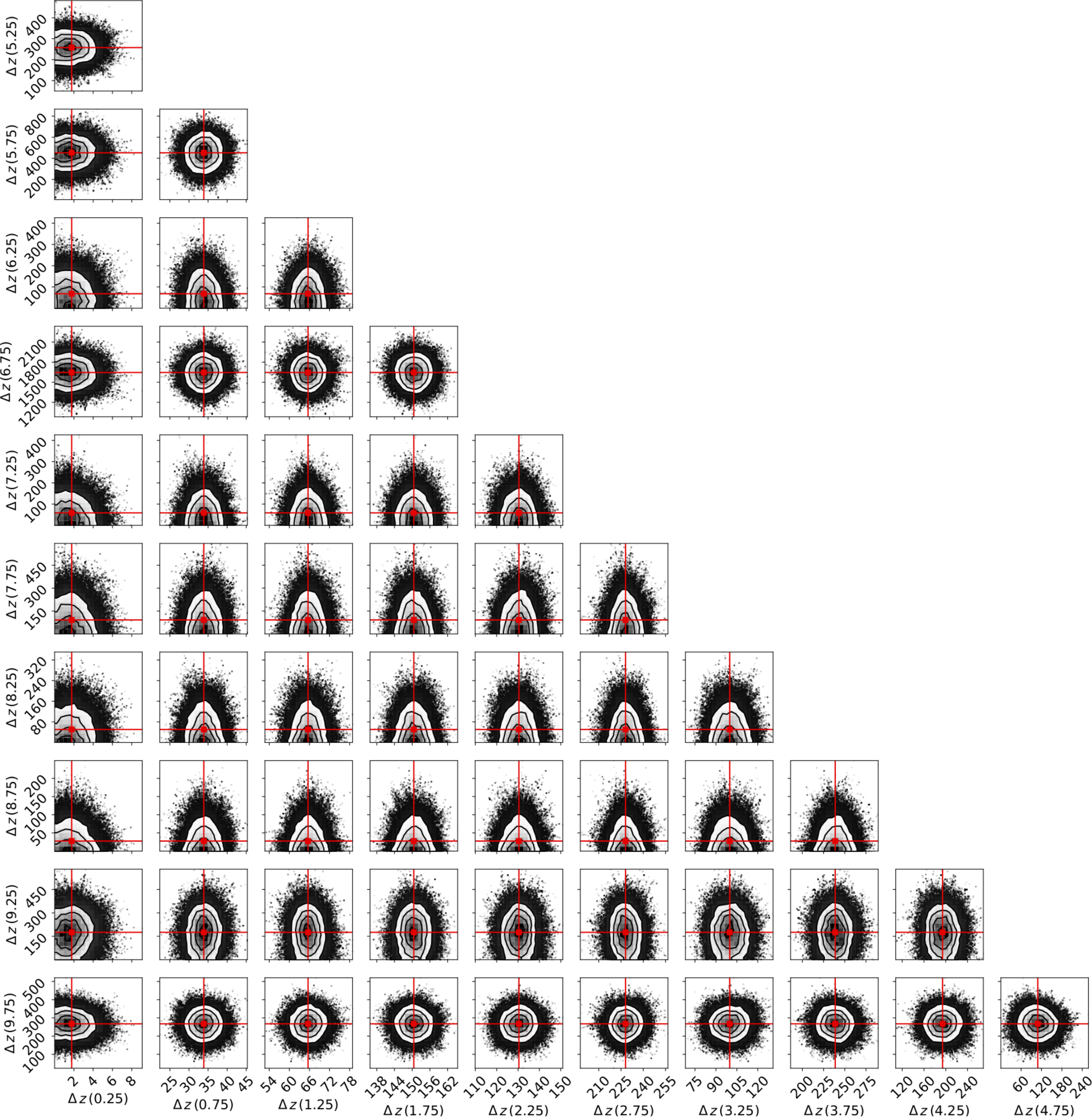}
\caption{Standard \texttt{emcee} corner plot showing the two-dimensional posterior distributions for our parameters (the average observed flux density in $\mu$Jy of galaxies in each redshift bin) for redshift intervals $\Delta z\,(0.25)$--$\Delta{z}\,(4.75)$ and $\Delta z\,(5.25)$--$\Delta{z}\,(9.75)$. Detailed description as in Figure \ref{fig:emcee_a}.  As evidenced by this plot there are only very weak correlations between our parameters for the redshifts intervals $\Delta z\,(0.25)$--$\Delta{z}\,(4.75)$ and $\Delta z\,(5.25)$--$\Delta{z}\,(9.75)$.}
\label{fig:emcee_c}
\end{figure*}

We aim to find the optimal set of average flux densities that minimizes the square of the residual flux between the model in equation \ref{eq conv M} and the observed beam-convolved map, weighted by the noise. We use the Markov chain Monte Carlo (MCMC) sampler \texttt{emcee} \citep{foreman13} to estimate the best fit flux densities and their uncertainties. We minimize a negative log likelihood $\ln(\mathcal{L})=-0.5\chi^2$, with
\begin{align}
\chi^2 = \sum_{ij}\left(\frac{\mathcal{O}_{ij}-\mathcal{M}_{ij}}{\sigma_{{\rm rms},{ij}}}\right)^2
\label{eq chi2}
\end{align}
where $\mathcal{O}$ is the observed map and $\sigma_{\rm rms}$ is the
instrumental noise map. We initialise $1000$ ``walkers'' with an
uninformative prior, such that each walker is set with a vector of
flux densities (representing $S_\nu(z)$) with each flux density drawn
from a Gaussian distribution of mean $0.5$\,mJy and width $0.05$\,mJy. The sampler runs for $1000$ iterations with the first $500$ iterations
discarded as burn-in. The best fitting flux densities and the
$1 \sigma$ bounds are estimated from the $16{\rm th}$, $50{\rm th}$ and $84{\rm th}$ percentiles of accepted samples for $500$ iterations.  In Figures \ref{fig:emcee_a}, \ref{fig:emcee_b} and \ref{fig:emcee_c} we show the \texttt{emcee} corner plot of the posterior distributions for all our free parameters.

To estimate the additional uncertainty on the stacked flux densities due to sampling variance we employ the ``delete one'' jackknife technique \citep{Tukey58}, splitting the map into $A=21$ approximately equal area sectors and running the MCMC fit for each jackknife.  We find that the sampler chains converge quickly (within $200$ steps), and tests indicate that the best fit parameters are insensitive to the initialisation parameters.  The covariance matrix is given by

\begin{align}
\mathcal{C}_{ij} = \frac{A-1}{A}\sum_{i=1}^A\left( S^k_{i} - \bar{S}_i\right)\left(S^k_j - \bar{S}_j\right)
\label{eq uncertainty}
\end{align}

where $S^k_i$ is the average flux density in the $i$th redshift bin, eliminating the $k$th sample and $\bar{S}_{i}$ is the average over all samples. The $1\sigma$ uncertainties on the stacked fluxes are estimated by the square root of the diagonal elements of $\mathcal{C}$. 

At high-$z$ the increase in the cosmic microwave background (CMB) temperature affects the measurement of submillimeter dust continuum in two ways \citep[see][for a detailed discussion]{da_cunha2013}.  Firstly, the CMB provides an additional source of dust heating increasing the intrinsic dust temperature as shown in equation \ref{eq dust_heating} \citep{da_cunha2013}: 
\begin{align}
T_{\mathrm {dust}}(z) = \Big( (T^{z=0}_{\mathrm {dust}})^{\beta +4} + (T^{z=0}_{\mathrm {CMB}})^{\beta+4}[(1+z)^{\beta+4}-1]\Big) ^{\frac{1}{\beta+4}}
\label{eq dust_heating}
\end{align}

Secondly, submillimeter observations of dust emission are always measured against the background of the CMB. At low-z $T_{\rm dust}(z) >> T_{\rm cmb}(z)$, so essentially all the intrinsic flux is detected against the CMB.  However, at high-z, as $T_{\rm cmb}(z)$ approaches $T_{\rm dust}(z)$, the fraction of submillimeter flux detected against the CMB decreases.  Equation \ref{eq fraction} \citep{da_cunha2013} shows the fraction of the intrinsic dust emission from a galaxy measured at a given frequency, $v_{\mathrm {obs}} = v_{\mathrm {rest}}/(1+z)$ against the CMB background: 
\begin{align}
\frac{F^{\mathrm{\, obs\, against\, CMB}}_{v_{\mathrm {obs}}}}{F^{\mathrm{intrinsic}}_{v_{\mathrm {obs}}}} = 1-\frac{B_v[T_{\mathrm {CMB}}(z)]}{B_v[T_{\mathrm {dust}}(z)]}
\label{eq fraction}
\end{align}
Assuming $T^{z=0}_{\mathrm {CMB}}=2.73\,$K, $T^{z=0}_{\mathrm {dust}}=25\,$K \citep[in line with the $T_{\mathrm {dust}}$ adopted in][]{Scoville_2016} and $\beta=2$ we derive the fraction of submillimeter flux observed against the CMB for the redshift range $0<z<10$ at $v_{\mathrm {obs}}{=}353$\,GHz ($\lambda_{\mathrm {obs}} {=} 850\,\mu$m), including the extra heating contributed by the CMB.  We apply this correction to our average observed $850\,\mu$m flux densities in all redshift bins to account for the impact of the CMB on our estimates.   

\section{Results}\label{section: results}
\subsection{Estimating molecular gas mass: RJ luminosity-to-gas mass relation}

In Table \ref{table:results1} we present the average observed $850\,\mu$m flux densities for our galaxy sample as a function of redshift. We quote the uncertainties to $1\sigma$ and include the additional uncertainty due to sample variance.  We note that at $z\gtrsim6$ the UDS redshifts are untested.  However, as our sample is binned according to the $\mathcal{P}(z)$ for each source, every galaxy effectively contributes to the flux in each redshift interval.  Hence, we show the average $850\,\mu$m flux density estimates for our galaxy sample to $z=10$.  

\begin{figure}
\figurenum{2}  
\includegraphics[width=0.48\textwidth]{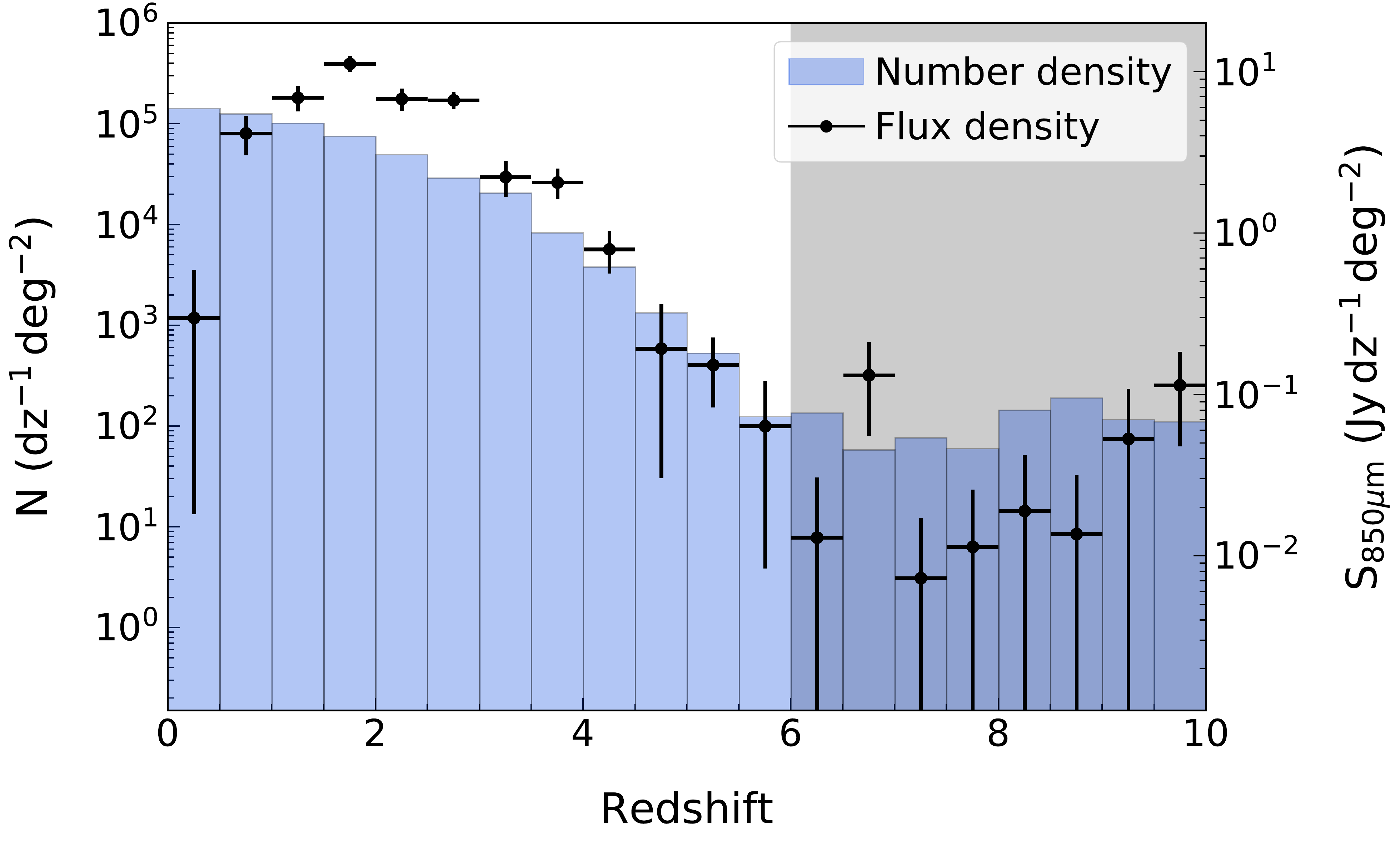}
\caption{Number density of galaxies in each redshift bin (blue bars) calculated from the galaxy ``weighting'' (the sum of the completeness corrected $\mathcal{P}(z)$ in each redshift bin) and the area of the unmasked region of the SCUBA-$2$ image.  The black points show the summed flux density of galaxies as a function of redshift derived by combining the number density and the average flux density of galaxies (see Table \ref{table:results1} column $4$ for average flux density values) in each redshift bin.  This figure shows the contrasting distributions, with the summed flux density mirroring the redshift distribution found for SMGs with a peak at $z\approx2$, whereas the number density generally shows a steady decline.  At $z\approx7$ there is an unexpected upturn in the number density and the summed flux density of galaxies, this is likely due to a combination of uncertainties in the photometric redshift fitting at high-$z$ (UDS photometric redshifts are untested at $z\gtrsim6$) and the small number statistics in these bins with the galaxy weighting being $\lesssim100$ per redshift interval.}  
\label{fig2:density}
\end{figure}

We sum $\mathcal{P}(z)$ (which is completeness corrected) in each redshift bin, giving us the galaxy ``weighting'' for each redshift interval. The integral of the summed $\mathcal{P}(z)$ across all redshift intervals should be approximately equal to the total number of galaxies in our sample. We calculate this to be $154,839$, which is consistent with our galaxy sample of $153,399$ sources (taking into account the completeness corrections).  With the summed $\mathcal{P}(z)$ and taking the area of our sample as the unmasked region of the SCUBA-$2$ $850\,\mu$m map (which corresponds to the UDS binary mask for ``good galaxy'' regions) we calculate the number density of galaxies as a function of redshift.  By combining this with our average flux density (see Table \ref{table:results1} column $4$ for CMB corrected values) we calculate the summed flux density for our galaxy sample in each redshift interval.  In Figure \ref{fig2:density} we present the number density and summed flux density for galaxies in our sample as a function of redshift.  The distribution of the summed flux densities with redshift is broadly comparable to the redshift distribution found for SMGs, which peaks at $z\approx2$ \citep[e.g.,][]{blain02,chapman05,simpson14,miettinen17,zavala18}, whilst the number density distribution generally declines with increasing redshift as expected.  The difference in the evolution of these distributions demonstrates that our derived flux densities are not biased by the number density of galaxies in each redshift bin.

\begin{deluxetable}{cccc}
\tablecaption{Redshift intervals, galaxy ``weighting'' in each redshift interval, average 850\,$\mu$m flux density, and average 850\,$\mu$m CMB corrected flux density. \label{table:results1}} 
\tablehead{
\colhead{\hspace{0.0cm}$\Delta z$}\hspace{0.02cm} & \colhead{\hspace{0.02cm}Galaxy weighting}\hspace{0.02cm} & \colhead{\hspace{0.02cm}$\langle S_{850}\rangle$}\hspace{0.16cm} & \colhead{\hspace{0.16cm}CMB corr. $\langle S_{850}\rangle$}\hspace{0.0cm}\\
& & \colhead{[$\mu$Jy\,dz$^{-1}$]}&\colhead{[$\mu$Jy\,dz$^{-1}$]}}
\startdata $0.25$ & $90003$ & $2.1\pm2.0$ & $2.1\pm2.0$ \\ [2pt]
$0.75$ & $79917$ & $32.9\pm8.2$ & $33.0\pm8.2$ \\ [2pt]
$1.25$ & $64475$ & $67.5\pm10.3$ & $68.0\pm10.3$ \\ [2pt]
$1.75$ & $47922$ & $146.9\pm12.8$ & $148.5\pm12.9$ \\ [2pt]
$2.25$ & $31448$ & $135.0\pm17.9$ & $137.2\pm18.2$ \\ [2pt]
$2.75$ & $18368$ & $224.9\pm21.5$ & $230.3\pm22.0$ \\ [2pt]
$3.25$ & $13071$ & $104.6\pm23.3$ & $108.3\pm24.1$ \\ [2pt]
$3.75$ & $5271$ & $236.5\pm45.1$ & $248.4\pm47.4$ \\ [2pt]
$4.25$ & $2413$ & $195.0\pm52.9$ & $209.1\pm56.7$ \\ [2pt]
$4.75$ & $848$ & $130.8\pm109.6$ & $144.3\pm120.9$ \\ [2pt]
$5.25$ & $336$ & $251.2\pm109.8$ & $288.4\pm126.1$ \\ [2pt]
$5.75$ & $79$ & $422.9\pm362.7$ & $511.9\pm439.0$ \\ [2pt]
$6.25$ & $86$ & $74.2\pm95.7$ & $96.2\pm124.1$ \\ [2pt]
$6.75$ & $37$ & $1606.2\pm868.8$ & $2276.3\pm1231.3$ \\ [2pt]
$7.25$ & $49$ & $60.2\pm77.3$ & $95.2\pm122.1$ \\ [2pt]
$7.75$ & $38$ & $106.0\pm127.0$ & $190.8\pm228.6$ \\ [2pt]
$8.25$ & $91$ & $63.2\pm73.7$ & $132.2\pm154.1$ \\ [2pt]
$8.75$ & $121$ & $29.0\pm36.5$ & $71.8\pm90.3$ \\ [2pt]
$9.25$ & $73$ & $155.2\pm152.5$ & $460.6\pm452.5$ \\ [2pt]
$9.75$ & $70$ & $289.5\pm161.3$ & $1039.0\pm578.9$ \\ [2pt]
\enddata
\tablecomments{Col.\ 1. Mid-point of each redshift interval ($\Delta z$); col.\ 2. the galaxy ``weighting'' which is the sum of the completeness corrected $\mathcal{P}(z)$ in each redshift bin. The summed $\mathcal{P}(z)$ across all redshift intervals integrates to $154,839$, which is consistent with the number of galaxies in our sample ($153,399$) taking into account the completeness corrections; col.\ 3. average (stacked) $850\,\mu$m flux density as a function of redshift with the uncertainty quoted to $1\sigma$; col.\ 4. average $850\,\mu$m flux density corrected for the impact of the CMB as a function of redshift with the uncertainty quoted to $1\sigma$.}
\end{deluxetable}

We adopt the approach of \citet[][hereafter \citetalias{Scoville_2016}]{Scoville_2016} and, utilising our flux density measurements, estimate the average molecular gas mass for our galaxy sample in each redshift interval. The full details of this approach are given in \citetalias{Scoville_2016}; however, we provide a brief description here. 

The long wavelength RJ tail of dust emission is nearly always optically thin ($\tau \ll 1$) and consequently this provides a direct probe of the total dust mass and hence the molecular gas mass.  \citetalias{Scoville_2016} utilise this to obtain an empirically calibrated \textit{RJ luminosity-to-gas mass} ratio

\begin{align}
\left(\frac{M_{\rm mol}}{\rm M_\odot}\right) = \frac{1}{\alpha_{850}}\left(\frac{L_{\rm 850, \ rest}}{\rm erg\,s^{-1}\,Hz^{-1}}\right) {\rm for}\ \lambda_{\rm rest}\gtrsim 250\,\mu{\rm m} 
\label{eq Mmol}
\end{align}
with $\alpha_{850}=6.7\pm1.7 \times 10^{19}\,{\rm erg}\,{\rm s}^{-1}\,{\rm Hz}^{-1}\,{\rm M_{\odot}}^{-1}$.  The restriction $\lambda_{\rm rest}\gtrsim 250\,\mu{\rm m}$ is required to ensure that at an observed wavelength of $850\,\mu$m the rest-frame emission stays on the RJ tail. \citetalias{Scoville_2016} demonstrate that this luminosity-to-mass ratio is relatively constant for high-stellar mass ($M_{\rm stellar}=(2-40)\times10^{10}\,{\rm M}_{\odot}$) normal star-forming and star-bursting galaxies, both locally and at high-\textit{z}. 

\newpage
We estimate the average rest-frame $850\,\mu$m luminosity density of galaxies as a function of redshift using the average flux densities detailed  in Table \ref{table:results1}, assuming a mass-weighted dust temperature of $25$\,K \footnote{The \citetalias{Scoville_2016} calibration uses a mass-weighted temperature of $25$\,K, rather than a luminosity-weighted dust temperature (see Appendix A.$2$ of \citetalias{Scoville_2016})} and employing the relation from \citetalias{Scoville_2016}:
\begin{multline}
L_{\nu 850} = S_\nu[\rm Jy] \ \times \ 1.19\ \times 10^{27} \ \times \ \left(\frac{\nu(850\,\mu{\rm m})}{\nu_{obs}(1+z)}\right)^{3.8} \\ \times \ \frac{(d_{\rm L}[{\rm Mpc}])^2}{1+z} \
\times \ \frac{\Gamma_{\rm RJ}(25,\nu_{850\,\mu{\rm m}},0)}{\Gamma_{\rm RJ}(25,\nu_{\rm obs},z)} \ [{\rm erg\,s^{-1}\,{\rm Hz}^{-1}}]
\label{eq:scoville_L}
\end{multline}

The $\Gamma_{\rm RJ}$ term in equation \ref{eq:rj} corrects for departures from the RJ $\nu^2$ dependence as the observed emission approaches the spectral energy distribution (SED) peak in the rest frame and where $T_{\rm dust}$ is the mass-weighted temperature characterizing the RJ dust emission
\begin{align}
\Gamma_{\rm RJ}(T_{\rm dust},\nu_{\rm obs},z) = \frac{h\nu_{\rm obs}(1+z)/kT_{\rm dust}}{e^{h\nu_{obs}(1+z)/kT_{\rm dust}}-1}\label{eq:rj}
\end{align}

\begin{deluxetable*}{ccccc}
\tablecaption{Redshift intervals, average rest-frame $850\mu$m luminosity, average molecular gas mass, co-moving molecular gas mass density as a function of $z$.\label{table:results2}} 
\tablehead{
\colhead{\hspace{0.2cm}$\Delta z$}\hspace{0.2cm} & \colhead{\hspace{0.2cm}$\langle L_{\rm 850,\, rest}\rangle$}\hspace{0.2cm} & \colhead{\hspace{0.2cm}$\langle M_{\rm H_2}\rangle$}\hspace{0.2cm} & \colhead{\hspace{0.2cm}$\rho_{\rm H_2}$}\hspace{0.2cm} & \colhead{\hspace{0.2cm}$\Omega_{\rm H_2}$}\hspace{0.2cm} \\
& \colhead{\hspace{0.2cm}[$10^{27}$\,erg\,s$^{-1 }$\,Hz$^{-1}$\,dz$^{-1}$]}\hspace{0.2cm} & \colhead{\hspace{0.2cm}[$10^7\,\rm {\rm M_\odot}$\,dz$^{-1}$]}\hspace{0.2cm} & \colhead{\hspace{0.2cm}[$10^6\,{\rm M_\odot}\,{\rm Mpc}^{-3}$]}\hspace{0.2cm} & \colhead{\hspace{0.2cm}[$10^{-7}$]}\hspace{0.2cm}}
\startdata $0.25$ & $1.64\pm0.02$ &  $1.80\pm0.03$ & $1.92\pm1.80$ & $115.66\pm108.42$ \\ [2pt]
$0.75$ & $81.97\pm3.89$ &  $89.95\pm4.26$ & $17.12\pm4.39$ & $569.51\pm146.13$ \\ [2pt]
$1.25$ & $217.79\pm18.32$ &  $239.01\pm20.10$ & $23.22\pm4.09$ & $430.70\pm75.93$ \\ [2pt]
$1.75$ & $514.61\pm64.72$ &  $564.77\pm71.02$ & $34.67\pm5.32$ & $380.54\pm58.41$ \\ [2pt]
$2.25$ & $492.12\pm85.19$ &  $540.08\pm93.49$ & $20.74\pm4.60$ & $143.48\pm31.84$ \\ [2pt]
\enddata
\tablecomments{
Col.\ 1. Mid-point of each redshift interval ($\Delta z$); cols. 2 and  3. the average rest-frame $850\,\mu$m luminosity and average molecular gas mass for galaxies in each redshift interval; col.\ 4. molecular gas mass density as a function of redshift and col.\ 5. molecular gas mass density in terms of the critical mass density. We restrict our results to $z\lesssim2.5$ to ensure that the observed $850\,\mu$m dust emission is tracing the RJ tail.}
\end{deluxetable*}

We restrict our estimates of the average rest-frame $850\,\mu$m luminosity to $z\lesssim2.5$ to ensure that rest-frame emission stays on the RJ tail. With the average rest-frame $850\,\mu$m luminosity density derived (detailed in Table \ref{table:results2}) we use the \textit{RJ luminosity-to-gas mass} ratio from equation \ref{eq Mmol} to estimate the average molecular gas mass as a function of redshift to $z\approx2.5$.  This calibration includes a factor of 1.36 to account for the associated mass of heavy elements (mostly Helium at $8\%$ by number), so we correct our results by a factor $1/1.36$ ($M_{\rm mol}$) to obtain $M_{\rm H_2}$.  

Since the summed photometric redshift probability distributions inform us about the galaxy ``weighting'' in each redshift interval and the UDS binary mask for ``good'' regions gives us the unmasked area of the SCUBA-$2$ $850\,\mu$m map, we combine this information with the average molecular gas mass and differential co-moving volume element to estimate the co-moving volume density of molecular gas
\begin{align}
\rho_{\rm H_2} = \Omega \int_{z-\Delta z/2}^{z+\Delta z/2} N(z)\langle M_{\rm H_2}(z)\rangle\frac{dV}{dz d\Omega}dz.
\label{eq mol gass density}
\end{align}
\noindent We present our values for $\rho_{\rm H_2}(z)$ in Table \ref{table:results2} as a function of redshift, also giving this in terms of the critical mass density $\Omega_{\rm H_2}=\rho_{\rm  H_2}(z)/\rho_{\rm crit}(z)$.  We use a Monte Carlo analysis to calculate the uncertainties for our values of ${L_{\rm 850,\, rest}}$, $M_{\rm H_2}$ and $\rho_{\rm H_2}(z)$, first drawing random values for $S_\nu$ from a Gaussian distribution where the mean is the average flux density and width the uncertainty on the average flux density, and then drawing values for a mass-weighted $T_{\rm dust}$ from a Gaussian distribution with a mean of $25$\,K (corresponding to the constant $T_{\rm dust}$ assumed by \citetalias{Scoville_2016}), and width $3$\,K.  Observations \citepalias[e.g.,][]{Planck11,Scoville_2016} and simulations \citep[e.g.,][]{Liang18,Liang19} find that a mass-weighted $T_{\rm dust}$ shows little variation with galaxy $L_{\rm 850}$ or redshift \citep[cf.][]{behrens18}, and by utilising a temperature distribution with $\sigma = \pm 3$\,K we recognise this minimal variance in our uncertainty calculations.  We use these values to estimate ${L_{\rm 850,\, rest}}$, $M_{\rm H_2}$ and $\rho_{\rm H_2}(z)$ from equations \ref{eq:scoville_L}, \ref{eq Mmol}, and \ref{eq mol gass density} respectively for $1000$ runs, with the uncertainty being taken as the standard deviation across these trials.

We note that the galaxy sample we use to derive the results in Tables \ref{table:results1} and \ref{table:results2} includes all galaxies in the ``good galaxy'' subset of the UDS DR11 catalog, regardless of the reliability of photometric redshifts for individual sources.  If we apply a $\chi^2$ cut to exclude galaxies with the least reliable redshifts (omitting galaxies with a reduced $\chi^2$ value for the photometric redshift of $>10$) and repeat the process outlined in Sections \ref{sect:method} and \ref{section: results} above we find a less than $2\%$ variation in our results with estimates consistent with those in Table \ref{table:results2} within the uncertainties.

We plot $\rho_{\rm H_2}(z)$ in Figure \ref{fig3:results}, compared to direct CO line estimates from ASPECS \citep{decarli16,Decarli20}, COLDZ \citep{riechers19} and VLASPECS \citep{Riechers_2020}, as well as values derived using far-infrared and UV photometry \citep{berta13}.  We fit a function of the same form as the star-formation rate density function presented in \citet{madau14} to the log of our results, and derive the best fit parameters for our data using \texttt{emcee}.  This yields:
\begin{multline}
{\rm log_{10}} \left(\frac{ \rho_{\rm H_2}}{\rm M_\odot\, {\rm Mpc^{-3}}} \right) = \\ (6.59\pm0.30) \times 
\frac{(1+z)^{0.38\pm0.16}}{1+[(1+z)/5.57\pm1.69]^{1.78\pm0.61}}
\end{multline}
 which we plot in Figure \ref{fig3:results}.  Our results show a peak $\rho_{\rm H_2}(z)$ at $z\approx2$ mirroring existing constraints.  

\subsection{Deriving molecular gas mass density from the halo mass function.}

Using an alternative approach we derive $\rho_{\rm H_2}(z)$ from first principles using the halo mass function from \citet{murray13} and assuming a constant halo mass range of $10^{11.5}$--$10^{15}\, {\rm M_\odot}$.  We estimate the molecular gas mass density as a function of halo mass (for redshifts $0\leq z\leq 7$) using the stellar-halo mass ratio from \citet{moster13} and the ISM-stellar mass relation from \citet{Scoville_2017}.  The ISM-stellar mass relation is calibrated using a sample of high mass galaxies ($M_{\rm {stellar}}\gtrsim10^{10}\,{\rm M}_\odot$), therefore we adopt a halo mass range for which the corresponding stellar masses are comparable with the \citet{Scoville_2017} calibration sample. Integrating these estimates with respect to halo mass gives the total molecular gas density as a function of redshift, which we present in Figure \ref{fig3:results}.    

\subsection{Estimating molecular gas mass density using a ``constant efficiency'' model}

\begin{figure*}
\includegraphics[width=1\linewidth]{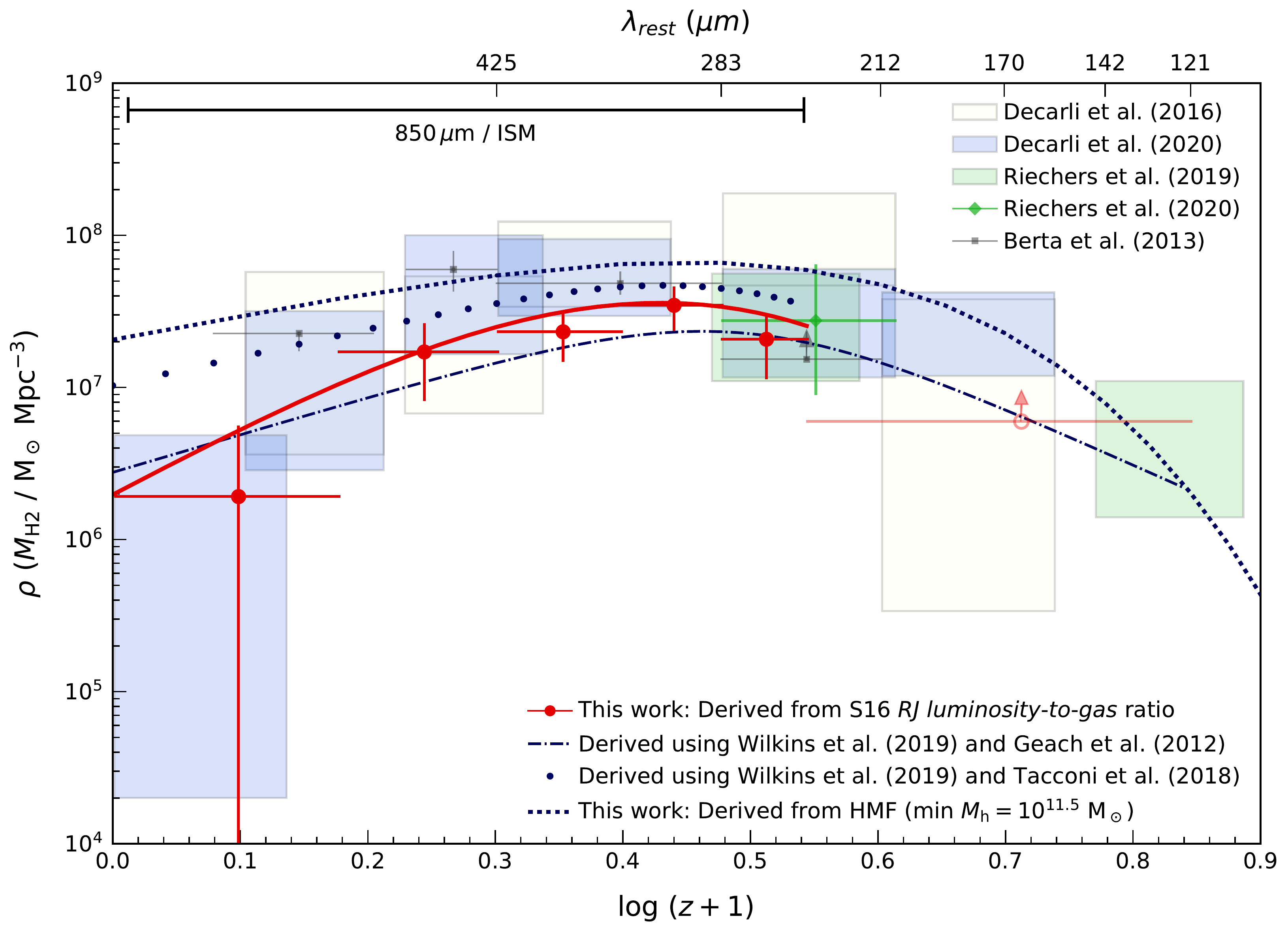}
\figurenum{3}
\caption{Values for $\rho_{\rm H_2}(z)$ (CMB corrected) derived using a $3$D stacking method and the \textit{RJ luminosity-to-gas mass} ratio of \citetalias{Scoville_2016}.  The upper x-axis shows the rest-frame wavelength of observed $850\,\mu$m emission for the redshift range shown, illustrating the range at which rest-frame emission traces the long-wavelength RJ tail (i.e. $\lambda_{\rm rest}\gtrsim250\mu$m) and the \citetalias{Scoville_2016} calibration can be reliably applied.  Our estimates are represented by the red points (uncertainties shown to $2\sigma$) with the solid red line showing the best-fit function derived using \texttt{emcee} and a function of the same form as the star-formation rate density function from \citet{madau14}.   We note that at $z\gtrsim2.5$ our results should be considered speculative as the observed $850\,\mu$m emission no longer traces the rest-frame RJ tail.  Hence we only show an average of our estimates after this point.  We do not include any estimates for $z\gtrsim6$ as the UDS photometric redshifts beyond this are untested and as such highly uncertain. Alongside our values we show results from ASPECS \citep[off white rectangles;][]{decarli16}, ASPECS LP \citep[blue rectangles;][]{Decarli20}, COLDz \citep[green rectangles;][]{riechers19} and VLASPECS \citep[green diamond;][]{Riechers_2020} which are derived from direct measurements of CO line emission.  The grey points show values from \citet{berta13}, estimated using deep far-infrared and UV data, and assuming either typical gas depletion times \citep{Tacconi13} or from IR luminosity and obscuration properties \citep{nordon13}. We plot  our ``constant efficiency'' models derived using the re-calibrated star formation history from \citet{wilkins19} and assuming either a corresponding constant \citep[e.g.,][]{geach12} or weakly evolving \citep[e.g.,][]{tacconi17} volume averaged star-formation ``efficiency'' to infer $\rho_{\rm H_2}(z)$.  The former is shown as the dark blue dot dash line and the latter illustrated by the dark blue circles.  We also plot $\rho_{\rm H_2}(z)$ to $z\approx7$ derived from the halo mass function \citep{murray13}, assuming the stellar-halo mass ratio from \citet{moster13} and ISM-stellar-mass relation from \citet{Scoville_2017}.  The dotted dark blue line corresponds to a halo mass range $10^{11.5}$--$10^{15}\,{\rm M_\odot}$, with the minimum stellar masses derived for this range (${\approx}10^{9.5}\,{\rm M_\odot}$) being consistent with the lowest stellar masses probed in ASPECS LP \citep{boogaard19}.}
\label{fig3:results}
\end{figure*}

We also estimate $\rho_{\rm H_2}(z)$ from the star-formation rate density, $\rho_{\dot{M_\star}}(z)$, assuming a corresponding volume averaged star-formation ``efficiency'', $\eta(z) = \rho_{\rm  H_2}(z)/\psi_\star(z)$.  We use the functional fit of \citet{wilkins19}, a recalibration of the well-known \citet{madau14} cosmic star formation history. We make the assumption that $\eta(z)$ is constant and that the {\it total} molecular gas mass per galaxy can be related to on-going star formation as $\xi M_{\rm H_2}={\rm SFR}/\epsilon$ \citep{geach12}. Here $\xi$ is the ratio of dense, actively star-forming molecular gas to the total molecular reservoir with $\xi\approx0.04$ for quiescent disks and $\xi>0.5$ for starbursts \citep[e.g.,][]{pap12}, while the factor $\epsilon$ describes the rate at which the dense molecular gas forms stars.  Figure \ref{fig3:results} shows the
predicted $\rho_{\rm H_2}(z)$ inferred from the \citet{wilkins19} fit, assuming a constant ``average'' $\eta(z)=0.3$\,Gyr corresponding to $\xi=0.1$ and $\epsilon=37$\,Gyr$^{-1}$ {\citep[e.g.,][]{geach12}}. 

This value for $\eta(z)$ is similar to the typical values of $t_{\rm dep}$ ($M_{\mathrm{H}_2}/\mathrm{SFR}$) $\approx1$\,Gyr \citep[e.g.,][]{tacconi17} for main-sequence galaxies.  \citet{tacconi17} find a relatively weak dependence of $t_{\rm dep}$ with redshift, $t_{\rm dep}\propto (1+z)^{-0.57}$  to $z<2.5$, consistent with our picture of a common mode of star formation in normal galaxies, at least out to the peak epoch.  We use this relation from \citet{tacconi17} and the \citet{wilkins19} fit to derive an estimate of $\rho_{\rm H_2}(z)$ which incorporates a weakly evolving star-formation efficiency.  We present our predicted $\rho_{\rm H_2}(z)$ in Figure \ref{fig3:results}.

\section{Discussion}

Our results appear to be in reasonable agreement with existing empirical constraints, indicating that the epoch of molecular gas coincided with the peak epoch of star formation at $z\approx2$ .  So what does this mean in terms of the evolving molecular gas budget?  We might ask what is the {\it complete} picture of $\rho_{\rm H_2}(z)$, or rather, what galaxies host the majority of the cosmic molecular gas budget across cosmic time? In the following discussion we interpret the overall
evolution of the cosmic molecular gas density, in the context of our results, within the established framework of star formation in galaxies from the cosmic dawn to the present day.

\subsection{Evolution of cosmic molecular gas mass density at \texorpdfstring{$0\leq z \leq 2.5$}{0< z < 2.5}}\label{results} 
The \citetalias{Scoville_2016} \textit{RJ luminosity-to-gas mass} ratio has been shown to provide molecular gas mass estimates accurate to within a factor of around $2$ when compared with measurements made via direct CO \,($J=1\rightarrow0$) line observations \citep[e.g.,][]{Scoville_2017,kaasinen19}, with variations in the dust emissivity index, temperature, and gas-to-dust ratios being accountable for the deviations.  This factor of $2$ accuracy is based on samples of galaxies with high stellar masses $M_{\rm stellar} = (2-40) \times 10^{10}\,{\rm M_\odot}$ as these galaxies are likely to have near-solar metallicity \citep{tremonti}.  This avoids probing low metallicity sources for which the dust-to-gas abundance ratio is likely to drop or the CO gas fraction is low \citep{bolatto13}.

\begin{figure}
\includegraphics[width=0.48\textwidth]{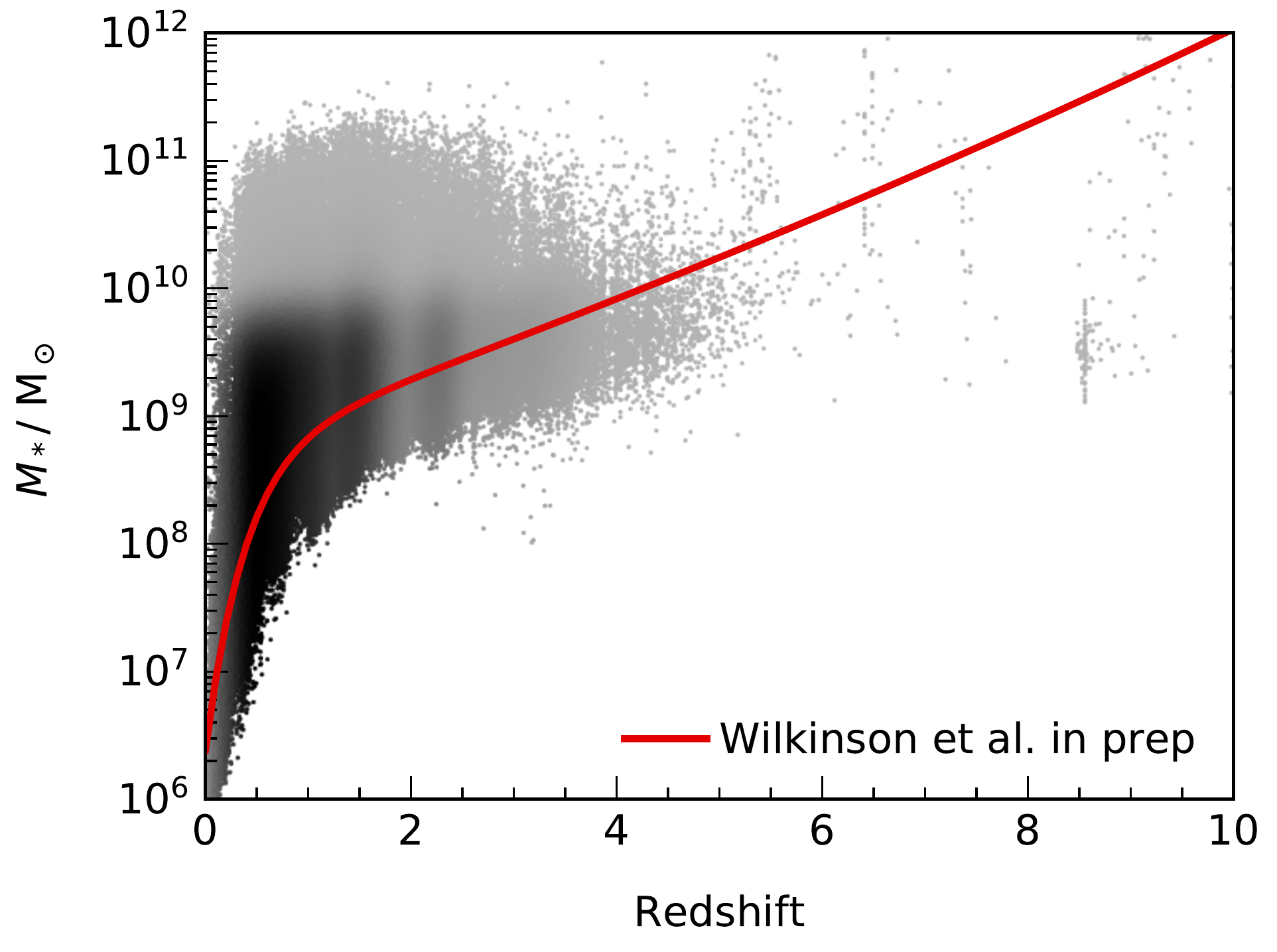}
\figurenum{4}
\caption{Marginalised stellar mass estimates (Almaini et al. in prep.) for UDS galaxies in our sample as a density plot, with darker colours corresponding to higher number densities of galaxies.  The red line corresponds to the UDS catalog $95\%$ stellar mass completeness \citep[derived using the method of][see Wilkinson et al. in prep.]{pozzetti2010}.  This figure demonstrates that at high redshifts we are only sensitive to the most massive galaxies.}
\label{fig4:klim}
\end{figure}

In Figure \ref{fig4:klim} we show the marginalised stellar mass estimates for UDS galaxies (Almaini et al. in prep.) and the corresponding $95\%$ stellar mass completeness \citep[derived using the method of][Wilkinson et al. in prep.]{pozzetti2010} for the UDS catalog.  As can be seen in Figure \ref{fig4:klim} our galaxy sample includes a proportion of galaxies with stellar masses lower than those used to derive the \citetalias{Scoville_2016} \textit{RJ luminosity-to-gas mass} ratio, with these sources being more abundant in lower redshift bins.

The dust-to-gas relation has been found to be relatively consistent for nearby galaxies with $M_{\rm stellar} > 10^{9}\, {\rm M_\odot}$ \citep[e.g.,][]{Groves15}, but drops for galaxies with lower stellar masses (hence lower metallicities).  Cosmological galaxy formation simulations have shown that deviations from this relation become significant ($\gtrsim 0.5\, {\rm dex}$) at $L_{\rm 850} \lesssim 10^{28}$\,erg\,s$^{-1}$\,Hz$^{-1}$ in the redshift range $0<z<9.5$ \citep[e.g.,][]{privon18}. 
 As shown in Figure \ref{fig4:klim}  at $z\gtrsim 1$ the majority of our sample are likely to have $M_{\rm stellar} > 10^{9}\, {\rm M_\odot}$ and Table \ref{table:results2} shows that the mean rest-frame $850\,\mu$m luminosity for our sample in all but the lowest redshift interval ($\Delta z=0.25$) is $L_{\rm 850\mu m} \gtrsim 10^{28}$\,erg\,s$^{-1}$\,Hz$^{-1}$. Therefore, whilst the \textit{RJ luminosity-to-gas mass} ratio has been calibrated on high stellar mass galaxies ($M_{\rm stellar} = (2-40) \times 10^{10}\,{\rm M_\odot}$), we make the assumption that applying this calibration to our sample at redshifts $z\gtrsim 0.5$ is likely to result in comparable uncertainties (i.e. a factor of $2$).  However, in the redshift bin $\Delta z=0.25$ the mean rest-frame luminosity is $\langle L_{\rm 850\mu m}\rangle = 1.6 \times 10^{27}$\,erg\,s$^{-1}$\,Hz$^{-1}$, and as such our results are likely to be under-predicted by $\gtrsim0.5$\,dex \citep[e.g.,][]{privon18} due to the abundance of lower mass (low metallicity) galaxies in this redshift interval.  

\subsection{Comparison of the evolution of molecular gas mass density to other studies in the literature}

In Figure \ref{fig3:results} we compare our results, which are revised to account for the influence of the CMB, to those from direct CO line surveys \citep[e.g.,][]{decarli16,Decarli20,riechers19,Riechers_2020}. We limit our discussion to the results from these surveys at $z\lesssim 2.5$ as we are restricted to this redshift range due to the wavelength of our observations ($\lambda_{\mathrm{obs}}=850\,\mu \mathrm{m}$).  Whilst the results from ASPECS/ASPECS LP \citep{decarli16,Decarli20} and COLDz \citep{riechers19,Riechers_2020} were not corrected for the influence of the CMB, we note that this is not necessary at $z\lesssim4.5$ as the effect of the CMB on measurements of molecular gas mass density from direct CO line observations is minimal \citep[$\lesssim 15\%$, e.g.,][]{Decarli19} and as such this does not impact on our analysis here.   
 
We find that our results are broadly consistent with the estimates from direct CO line surveys within uncertainties and show notably good agreement with results obtained through observations of the ground state CO line (ASPECS LP at $\Delta z \simeq 0.25$ and COLDZ at $\Delta z \simeq 2.25$). Albeit, we caution that our results at $\Delta z \simeq 0.25$ are likely under estimated due to the abundance of low stellar mass galaxies in this redshift bin.  

When compared with the ASPECS LP survey \citep{Decarli20} our results generally trace the lower boundaries of their estimates between $0.75\lesssim z\lesssim1.75$.  The \citetalias{Scoville_2016} \textit{RJ luminosity-to-gas mass} ratio is calibrated using the ground state CO\,(${J=1\rightarrow0}$) line, whereas at $z>0.75$ the ASPECS LP results are derived from observations of higher state excitation CO lines.  Therefore, this offset could be explained by the uncertainties associated with translating higher excitation CO lines observations to ground state CO\,(${J=1\rightarrow0}$) luminosities.  However, in the redshift interval $\Delta z=1.25$, even if the extreme case of thermalized gas is assumed,  our upper estimate (taking into account $1\sigma$ uncertainties) falls a factor of $1.42$ below the lower boundary of the ASPECS LP survey \citep[e.g., $38.90\times10^6\,{\rm M_\odot}\,{\rm Mpc}^{-3}$, Table A3.][]{Decarli19}.  As such the uncertainties in CO line ratios do not fully account for the offset we see.   

Building on previous studies \citet{liu2019} derive the molecular gas mass density using a dataset comprised of ${\approx}700$ ALMA continuum detected galaxies and ${\approx} 1000$ galaxies with CO observations (taken from the literature). To derive molecular gas masses for the continuum detected galaxies \citet{liu2019} employ the \citet{hughes17} \textit{luminosity-to-gas mass} calibration.  \citet{liu2019} estimate a SMF (stellar mass function) integrated molecular gas mass density based on the SMF integrated to $M_{\rm stellar} = 10^9\,{\rm M_\odot}$ and using a gas fraction function derived from their composite sample of ${\approx}1700$ galaxies.  Their results trace the upper boundaries of the molecular gas mass density derived from the most recent blank field CO line surveys \citep[e.g.,][]{Decarli19,riechers19} and are ${\approx}1$\,dex higher than our estimates.  This offset with our estimates of molecular gas mass density could be in part due to assumptions made in \citet{liu2019} to derive an SMF integrated molecular gas mass density (i.e. that all star-forming galaxies are on the main-sequence) or potentially differences in sample selection.  The majority (${\approx} 800$) of the CO detected sources in the \citet{liu2019} composite sample are in the Local Universe (i.e $z<0.3$), so at $z>0.3$ their dataset is dominated by ALMA continuum detected galaxies which are preferentially massive and dust-rich (and hence, using a dust-to-gas mass conversion, gas-rich).  In contrast our galaxy sample is near-infrared selected and as such our selection is less likely to sample these luminous dust-rich SMGs, with previous studies finding that $\approx{20}\%$ of SMGs are missed in optical/new-infrared surveys \citep[e.g.][]{dudzevi19}.

\citet{magnelli20} use a stacking approach to measure the comoving gas mass density of a sample of $555$ near-infrared selected galaxies, with galaxies split into bins of $z$ and ${M}_{\rm stellar}$.  Their stacking method accounts for the metallicity of galaxies in these bins \citep[inferred using the stellar mass-metallicity from][]{tacconi17} and is equivalent to the \citetalias{Scoville_2016} calibration at solar metallicity.  Our results trace the lower boundaries of their estimates, but are inconsistent (within $1\sigma$ uncertainties) at $z>1$.  This discrepancy may be in part due to our method not accounting for the metallicity of low mass galaxies, resulting in an under-estimation of the molecular gas mass for galaxies with $M_{\rm stellar} < 10^{9}\, {\rm M_\odot}$.  However, if this was the sole reason for the difference in our results we would expect this to have more of an impact at $z<1$ where this effect will be more prominent.

We caution that our results also rely solely on photometric redshifts, which despite the high quality 12 band photometry of the UDS catalog, cannot compete with the accuracy of redshifts derived via spectroscopic surveys.  By utilising the redshift probability distributions in our 3D stacking approach we aim to provide mitigation against these uncertainties.  However, whilst our estimates rely exclusively on the use of photometric redshifts, the results obtained in both ASPECS LP \citep{Decarli20,magnelli20} and \citet{liu2019} benefit from the inclusion of sources with more reliable spectroscopic redshifts.  This may also play a part in deviations seen when we compare our estimates with these previous surveys.

\subsection{Contribution of the brightest submillimeter sources to the cosmic evolution of the molecular gas mass density}

In order to present the most complete view of the evolution of the molecular gas mass density we stack all sources in our near-infrared selected sample, including counterparts to the bright submillimeter sources in the SCUBA-$2$ UDS map.  To test the contribution of these galaxies to our results we repeat our stacking analysis with the SCUBA$2$ UDS source subtracted map. As expected excluding the $\sim1000$ UDS submillimeter sources reduces the average observed $850\,\mu$m flux in our redshift intervals, which propagates to our estimate of the comoving molecular gas mass density.  At $z<1.5$ the exclusion of the UDS submillimeter sources has a minimal impact on our estimates of $\rho_{\rm H_2}(z)$ and these remain consistent within the $1\sigma$ uncertainties.  However, at $z>1.5$ our estimates of $\rho_{\rm H_2}(z)$ drop by a factor of $2.05$ and $2.33$ in the redshift intervals $\Delta z = 1.75$ and $\Delta z = 2.25$ respectively.  This coincides with the peak number density of SMGs at $z\approx2$.  This indicates that approximately $50\%$ of the molecular gas mass density at the peak of the star formation rate density is locked in dust-rich SMGs.  We note that our inferred contribution of SMGs is also likely under-estimated as we expect that approximately $20\%$ of SMGs are undetected in our near-infrared selected sample \citep[e.g.,][]{dudzevi19}.  Our finding is in keeping with \citet{zavala21} who find that bright SMGS ($L_{\rm IR}>10^{12}L_\odot$) dominate the obscured star formation rate density at $z\approx2$ and also \citet{magnelli20} who find that the bulk of dust and gas in galaxies is locked in massive star-forming galaxies.

\subsection{Additional constraints on the evolution of molecular gas mass density} 

We have added further valuable constraints to this picture of cosmic molecular gas evolution using two alternative approaches.

We estimate $\rho_{\rm H_2}(z)$ from the halo mass function \citep{murray13} assuming a constant halo mass range of  $10^{11.5}$--$10^{15}\,{\rm M_\odot}$, and  using the stellar-halo mass ratio from \citet{moster13} and the ISM-stellar mass relation from \citet{Scoville_2017}. For the latter relation we make the assumption that all galaxies are on the star-forming main sequence (e.g., $\rm{sSFR/sSFR_{MS}}=1$). The evolution of our halo mass derived $\rho_{\rm H_2}(z)$(shown in Figure \ref{fig3:results}) follows a similar shape to the star-formation rate density, rising to a peak at $1\lesssim z \lesssim 3$ and decreasing to the present day. The minimum halo mass we assume corresponds to stellar masses of ${\approx}10^{9.5}\,{\rm M_\odot}$ \citep[e.g.,][]{moster13}, equivalent to the lowest stellar masses probed in the ASPECS LP survey \citep{Decarli20}.  Our $\rho_{\rm H_2}(z)$ estimates show good agreement with the ASPECs/ASPECS LP surveys \citep{decarli16b,Decarli20} at $z\gtrsim0.6$. However, as shown in Figure \ref{fig3:results} at $z\lesssim0.6$ our estimate of $\rho_{\rm H_2}(z)$ lies above the lowest redshift bins from the  ASPECS LP survey \citep{Decarli20} and is ${\approx}1\, \rm{dex}$ higher than our estimate of $\rho_{\rm H_2}(z)$ derived from measurements of observed $850\,\mu$m flux.  To obtain an estimate of $\rho_{\rm H_2}(z)$ from the halo mass function we make the assumption that all galaxies are star-forming.  As such our estimate of $\rho_{\rm H_2}(z)$ derived from the halo mass function can be seen as an upper limit $\rho_{\rm H_2}(z)$ for the stellar mass range sampled.  It follows that we see a more significant deviation between our estimate and observationally derived results at lower redshifts as the fraction of passive galaxies is higher at later epochs.              

    We also estimate $\rho_{\rm H_2}(z)$ from the star-formation rate density \citep{wilkins19}, assuming a constant \citep{geach12} and weakly evolving \citep{tacconi17} star-formation efficiency.  These ``constant efficiency'' models predict a co-moving molecular gas mass density in good agreement with both measurements of molecular gas mass via observations of direct CO line emission \citep{decarli16,Decarli20,riechers19,Riechers_2020} and our results derived from measurements of the long-wavelength dust emission, out to a peak at $z\approx2$. A simple conclusion is that the peak epoch of star formation at $z\approx2$ is not driven by significantly more efficient (or starburst-like) star formation in galaxies, but by a higher abundance of molecular fuel in galaxies. We note that the estimate derived from weakly evolving star-formation is ${\approx}1\,\rm{dex}$ higher than our results at $\Delta z \simeq 0.25$.  This is likely a consequence of the latter being under estimated due to the abundance of low stellar mass galaxies in this redshift bin.       

We recognise that our assumption of a ``constant efficiency'' model is at odds with \citet{Scoville_2017}, who argue that whilst cold molecular gas reservoirs increase with $z$ (as $(1+z)^{1.84}$), the star-formation rate increases more rapidly (as $(1+z)^{2.9}$), indicating that the peak of star formation is a consequence of both increased molecular gas content in galaxies and higher star-formation efficiency.  We also note that at $z\gtrsim 1$ early-type galaxies have been shown to be more compact for a given stellar mass than their local counterparts \citep[e.g.,][]{daddi05,cappellari09}, which taken in combination with the ``Kennicutt-Schmidt'' relation \citep[a power-law relation between star-formation rate and gas surface densities,][]{kennicutt98,schmidt59} implies that star formation may be more efficient at $z\gtrsim1$. 

The $3$D stacking approach we use derives the \textit{average} properties for galaxies in our sample as a function of redshift, and thus we do not measure the molecular gas mass and star-formation rates for individual sources.  Whilst the UDS DR11 catalog does include M$_{\rm stellar}$ estimates (which are evaluated at the peak maximum likelihood redshift) for individual galaxies, our $3$D stacking method bins galaxies according to the discretized redshift probability distribution ($\mathcal{P}(z)$), and as such each galaxy in our sample effectively contributes to the flux in all redshift intervals.  Hence, using this $3$D stacking method precludes a M$_{\rm stellar}$ selection relative to our redshift bins.  Therefore, we are not able to repeat the analysis from \citet{Scoville_2017} to test their assertion of an evolving star-formation efficiency.

\subsection{The epoch of molecular gas}

Although we cannot quantify the contribution of higher star-formation efficiencies to the peak of star-formation rate density at $z\approx2$, the symmetry between our ``constant efficiency'' models with our statistically derived $\rho_{\rm H_2}(z)$ indicates a star formation history which is predominantly driven by an increased supply of molecular gas in galaxies, rather than a significant evolution in star-formation efficiency \citep[consistent with the findings of][]{Decarli20,magnelli20}.  With this in mind we now turn to the formation of H$_2$ itself.

\cite{cazaux04} combine a microscopic model for the relative rates of gas-phase and dust H$_2$ production with a cosmological model to show the more efficient dust-phase production becomes the dominant route to H$_2$ formation at $z\approx3$--$6$ for reasonable assumptions about the conditions of the interstellar medium of early galaxies. 
Therefore, there is a perfect storm for massive galaxy growth at $z\approx2$: not only is the cosmic accretion rate at its peak, massive halos have had time to grow, galaxies have increased gas densities, and previous generations of stars in the progenitors of these systems have provided the metal enrichment that accelerates the formation of H$_2$, which, as the fuel for star formation, drives galaxy growth; this could be described as the epoch of molecular gas.

\subsection{Estimating the evolution of cosmic molecular gas mass density at \texorpdfstring{$z\gtrsim 2.5$}{z<6}}\label{spec results}

\citetalias{Scoville_2016} intentionally restrict their calibration sample to galaxies at $z\leq3$ to ensure observed $850\,\mu$m emission is from the rest-frame long wavelength RJ tail, where dust is optically thin and emission is dominated by the contribution of cold dust (which is well represented by a mass-weighted $T_{\rm dust}=25$\,K).  In Figure $3$ we have shown an average of our $\rho_{\rm H_2}(z)$ estimates at $2.5\lesssim z\lesssim6$ (the UDS redshifts are untested at earlier epochs as there are no UDS galaxies with spectroscopic redshifts at $z\gtrsim6.5$), but note that at these redshifts estimates are less reliable due to large uncertainties in the RJ correction (see equation \ref{eq:rj}) as rest-frame emission approaches the peak of the SED. 

In the optically thick regime (as rest-frame dust emission moves off the long-wavelength RJ tail) the rest-frame emission no longer correlates with the total dust mass of a galaxy and probes only the surface dust, which using the approach of \citetalias{Scoville_2016} would result in under-estimation of $L_{\rm 850}$ and hence the molecular gas mass.  However, as the rest-frame emission approaches the peak of the SED we are increasingly sensitive to the dense, warm dust component, which significantly boosts the luminosity (with only a small mass fraction) and dominates the emission close to the SED peak.  Consequently rest-frame dust emission at high-$z$ is not well represented by a mass-weighted $T_{\rm dust}=25$\,K, which would result in a over-estimate of the dust and gas mass.  

In addition to these competing effects, we are also likely to be missing a significant population of lower mass galaxies at $z>2.5$.  As shown in Figure \ref{fig4:klim} the $95\%$ stellar mass completeness at $z\approx2.5$ is predicted to be $\simeq10^{9.5}\, {\rm M_\odot}$, so we are are simply not sensitive to the majority of low mass galaxies at the highest redshifts.   In addition, although relatively rare \citep[with number counts $\mathrm{N}(>3.5\,\mathrm{mJy) \simeq 3000 \ deg}^{-2}$;][]{geach17} SMGs are dust-rich   \citep[$M_{\rm dust}\sim10^9\,{\rm M_\odot}$; e.g.,][]{da_cunha2015,magnelli19} and about $20\%$ are undetected in optical/near-infrared surveys \citep[e.g.,][]{dudzevi19}.  This non-detection of SMGs is unlikely to have a significant impact on our estimates at low $z$.  However, at $z\gtrsim2.5$ since we are significantly under-sampling the galaxy population and as the number of galaxies in our redshift bins fall the non-detection of dust-rich SMGs becomes more statistically significant, further contributing to an under-estimation of the molecular gas mass density at $z\gtrsim2.5$.

The overall impact of the above is difficult to quantify.  However, as shown by Figure \ref{fig3:results} our results at $z\gtrsim2.5$ are systematically lower than the estimates obtained via direct CO line emission, which suggests that the use of this method past $z\approx2.5$ (when $\lambda_{obs} = 850\,\mu$m  no longer probes the rest-frame RJ tail) results in an under-estimation of the molecular gas mass density.  In consequence, whilst our results are highly uncertain at $z>2.5$ we suggest that to $z\lesssim6$ these can be seen as providing a lower-limit to the molecular gas mass density.  

\section{CONCLUSIONS}
We employ a $3$-dimensional stacking method \citep{Viero13} and an empirically calibrated \textit{RJ luminosity-to-gas mass} ratio \citepalias{Scoville_2016} to derive the average molecular gas mass as a function of redshift utilising a sample of ${\approx} 150,000$ galaxies in the UKIDSS-UDS field. By combining these techniques we are able to reduce the statistical uncertainties on the evolution of the molecular gas mass density, $\rho_{\rm H_2}(z)$, within the redshift range $0.5\lesssim z \lesssim 2.5$.  We find that:
\smallskip

\textbullet{\ $\rho_{\rm H_2}(z)$ shows a clear evolution over cosmic time which traces that of the star-formation rate density with a peak $\approx2\times10^7\,{\rm M_\odot}\,{\rm Mpc^{-3}}$ at $z\approx2$.}  
\smallskip

\textbullet{\ Our results are consistent with those of blank field} CO line surveys, albeit our estimates are systematically lower than those derived using observations of higher excitation CO lines.  This may in part be a consequence of the line ratios used to translate higher excitation CO line luminosity to ground state CO line luminosity.
\smallskip

\textbullet{\ Our results are an order of magnitude lower than those derived by \citet{liu2019} who use the \citet{hughes17} \textit{luminosity-to-gas mass} calibration to estimate molecular gas masses for the ALMA continuum detected galaxies in their sample.  This difference in results may be in part due to selection effects, as their ALMA-selected sample preferentially selects dust-rich (and consequently gas-rich), sources, whereas by using a NIR selection we are likely to miss $\approx{20}\%$ of these dust-rich SMGs}.
\smallskip

\textbullet{\ $\rho_{\rm H_2}(z)$ can be broadly modelled by inverting the star-formation rate density \citep{wilkins19} with a constant \citep{geach12} or weakly evolving \citep{tacconi17} volume averaged star-formation efficiency.  Our ``constant efficiency'' models closely align to our statistically derived $\rho_{\rm H_2}(z)$.}  
\smallskip

\textbullet{\ $\rho_{\rm H_2}(z)$ can be derived from first principles from the halo mass function \citep{murray13} in conjunction with stellar-halo mass \citep{moster13} and ISM-stellar mass ratios \citep{Scoville_2017}.  To obtain this estimate we make the assumption that all galaxies are star-forming and hence this can be seen as an upper limit for $\rho_{\rm H_2}(z)$ with respect to the stellar mass range sampled.}
\smallskip

We have demonstrated that by applying a statistical method and the approach of \citet{Scoville_2016} we can provide robust, statistically significant constraints to the cosmological gas mass density to $z\lesssim2.5$.  Our results show an evolution that mirrors that of the star-formation rate density indicating that the peak of the star formation history is primarily driven by an increased supply of molecular gas rather than a significantly increased star-formation efficiency. We have shown that at $z\gtrsim2.5$ we detect dust emission from high mass galaxies, even with our near-infrared selected sample.  Hence, in the future there is potential for this approach to be extended to provide improved constraints at higher-$z$ through $1$\,mm/$3$\,mm wide-field surveys with facilities such as the Large Millimeter Telescope.

\acknowledgements
{We thank the anonymous  referee  for  their  helpful comments. TKG acknowledges support from a UK Science and Technology Facilities Council studentship. KEKC is supported by a Royal Society Leverhulme Senior Research Fellowship (SRF/R1/191013). KEKC and MF acknowledge the support from STFC (grant number ST/R000905/1).  JEG is supported by a Royal Society University Research Fellowship. CCL acknowledges support from the Royal Society under grant RGF/EA/181016.  MPK acknowledges support from the First TEAM grant of the Foundation for Polish Science No. POIR.04.04.00-00-5D21/18-00. MPK also acknowledges support from the Polish National Agency for Academic Exchange under Grant No. PPI/APM/2018/1/00036/U/001.  The James Clerk Maxwell Telescope has historically been operated by the Joint Astronomy Centre on behalf of the Science and Technology Facilities Council of the United Kingdom, the National Research Council of Canada and the Netherlands Organisation for Scientific Research.  The S2CLS map data were taken as part of Program ID MJLSC02. Additional funds for the construction of SCUBA-2 were provided by the Canada Foundation for Innovation. The UKIDSS project is defined in \citet{lawrence07}. Further details on the UDS can be found in Almaini et al. (in prep.). UKIDSS uses the UKIRT Wide Field Camera (WFCAM; \citealt{casali07}). The photometric system is described in \citet{hewett06}, and the calibration is described in \citet{hodgkin09}. The pipeline processing and science archive are described in Irwin et al (in prep.) and \citet{hambley08}.}

\bibliography{ref1.bib}
\bibliographystyle{aasjournal.bst}

\end{document}